\documentclass[prd,
               twocolumn,
               amssymb,
               amsmath,
               eqsecnum,
               showpacs,
               letterpaper,
               superscriptaddress,
               altaffilletter]
               {revtex4}

\usepackage{color}
\usepackage{graphicx}
\usepackage{latexsym}
\usepackage{float}
\usepackage{amsmath}
\usepackage{amssymb}
\usepackage{multirow}

\newcommand{\bfxi}{\boldsymbol\xi}

\begin{document}

\title{Method for detection and reconstruction of gravitational wave transients \\ 
with networks of advanced detectors. }

\author{S.~Klimenko}  
\affiliation{University of Florida, P.O.Box 118440, Gainesville, Florida, 32611, USA}
\author{G.~Vedovato}  
\affiliation{INFN, Sezione di Padova, via Marzolo 8, 35131 Padova, Italy}
\author{M.~Drago} 
\affiliation{Max Planck Institut f\"{u}r Gravitationsphysik, Callinstrasse 38, 
             30167 Hannover, and Leibniz Universit\"{a}t Hannover, Hannover, Germany}
\author{F.~Salemi} 
\affiliation{Max Planck Institut f\"{u}r Gravitationsphysik, Callinstrasse 38, 
             30167 Hannover, and Leibniz Universit\"{a}t Hannover, Hannover, Germany}
\author{V.~Tiwari}
\affiliation{University of Florida, P.O.Box 118440, Gainesville, Florida, 32611, USA}
\author{G.~A.~Prodi}    
\affiliation{University of Trento, Physics Department and INFN, Trento Institute 
for Fundamental Physics and Applications, via Sommarive 14, 38123 Povo, Trento, Italy}
\author{C.~Lazzaro}  
\affiliation{INFN, Sezione di Padova, via Marzolo 8, 35131 Padova, Italy}
\author{K.~Ackley} 
\affiliation{University of Florida, P.O.Box 118440, Gainesville, Florida, 32611, USA}
\author{S.~Tiwari}  
\affiliation{University of Trento, Physics Department and INFN, Trento Institute 
for Fundamental Physics and Applications, via Sommarive 14, 38123 Povo, Trento, Italy}
\affiliation{Gran Sasso Science Institute (INFN), Via F. Crispi 7, I-67100, L’Aquila, Italy}
\author{C.~F.~Da~Silva~Costa} 
\affiliation{University of Florida, P.O.Box 118440, Gainesville, Florida, 32611, USA}
\author{G.~Mitselmakher}
\affiliation{University of Florida, P.O.Box 118440, Gainesville, Florida, 32611, USA}


\begin{abstract}

We present a method for detection and reconstruction of the gravitational-wave (GW)
transients  with the networks of advanced detectors. Originally designed to 
search for transients with the initial GW detectors, it uses significantly improved algorithms, 
which enhances both the low-latency searches with 
rapid localization  of GW events for the electro-magnetic followup and high confidence 
detection of a broad range of the transient GW sources. In the paper we present the analytic 
framework of the method. Following a short description of the core analysis algorithms,
we introduce a novel approach to the reconstruction of the GW polarization
from a pattern of detector responses to a GW signal. This polarization pattern is 
a unique signature of an arbitrary GW signal that can be measured independent from 
the other source parameters. The polarization measurements enable rapid 
reconstruction of the GW waveforms, sky localization and helps  identification of 
the source origin.

\end{abstract}


\date[\relax]{Dated: \today }

\maketitle


\section{Introduction}

Advanced LIGO detectors~\cite{TheLIGOScientific:2014jea} has started their operation at unprecedented
sensitivity targeting first detection of gravitational waves 
from astrophysical sources. A more robust detection of gravitational waves
is anticipated in the next few years as the advanced LIGO 
reaches its designed sensitivity and the other advanced detectors 
Virgo~\cite{Virgo}, Kagra~\cite{kagra} and LIGO-India~\cite{LIGOIndia} come online. 
Numerous GW signals, expected to be observed by the advanced
detectors ($\sim{40}$ binary neutron star and possible black hole
mergers per year~\cite{ExpRates})
will begin our exploration of the gravitational-wave sky and start the era of 
the gravitational wave astronomy.  

The advanced detectors target detection of GW transients for a wide range of
promising astrophysical sources including: various types of gamma-ray bursts, 
core-collapse supernovae, soft-gamma repeaters, cosmic strings, late inspiral 
and mergers of compact binaries, ring-downs of perturbed neutron stars or black holes, 
and as-yet-unknown systems. Most of these sources are difficult to model, due 
to their complicated dynamics and because the equation of state of matter at neutron 
star densities is not known. Therefore, the search algorithms have been 
developed~\cite{klimenko:2008kymm,Xpipeline,STAMP,oLIB}
for detection of GW transients, or bursts of GW radiation in 
the detector bandwidth, with no 
or little assumptions on the source models.

There are two different ways the GW searches are conducted: in real time and 
searches on the archived data. 
The objective of the real time burst search is the identification and reconstruction of significant 
event candidates with low latency (within few minutes). The reconstructed sky location
can be promptly shared with the partner telescopes, which search for a coincident 
electromagnetic (EM) counterpart~\cite{Abbott:2011ys,Aasi:2013sna}. 
A prominent source for such join observation is a merger 
of compact binary objects where one of the companions (or both) is a neutron star. Such mergers 
may produce several EM signals: gamma-ray busts (GRB), GRB afterglow, kilonova, etc, 
which will fade away with the time scales ranging from seconds to days~\cite{kilonova:2012mb}. 
A small fraction of such mergers (when the GRB beam is pointing at us) can be independently 
detected by the gamma-ray telescopes and associated with a GW signal 
by the time of the event. However, most of the compact binary mergers require 
a prompt sky localization with the GW detectors and follow-up EM searches for possible afterglow. Similar observations can be performed for 
the galactic events such as supernovae or soft-gamma-repeaters, which may produce both 
the EM and neutrino counterparts. On contrary, the objective of the archived burst analysis is 
to establish a significance of observed events and identify their progenitors. 
Such analysis requires detail background studies and accurate reconstruction of 
the source parameters, which may not be readily available with low latency.

Both types of searches and the sky localization studies have been performed 
with the baseline burst algorithm coherent WaveBurst (cWB)~\cite{klimenko:2008kymm} 
used in the analysis of data form the initial 
instruments~\cite{Online1,IMBBH0,Burst1,IMBBH1,Burst2,IMBBH2,Klimenko:2011sky}.
In this paper we describe the improvements of the cWB algorithm, which is currently 
used both for the real time burst search and several archived searches with the networks 
of advanced detectors.
This second generation cWB algorithm  includes several novelties. 
The time-frequency analysis has been updated with a novel time frequency transform~\cite{WDM}, 
which improved the waveform reconstruction. 
It also significantly improved the computational performance of the algorithm,
enabling a robust low latency operation. 
The data conditioning (whitening, removal of the spectral artifacts, etc) has been 
enhanced with the data regression algorithms~\cite{Tiwari:2015tkm}. Fast 
reconstruction of the chirp mass~\cite{Tiwari:2015chirp} has been introduced to enable 
rapid identification of the compact binary coalescence (CBC) sources. 
The extensive sky localization studies have been performed~\cite{aLIGOskyloc}. 

In this paper we focus on the cWB analytic framework enhanced with a novel method for 
reconstruction of the GW polarization from the pattern of detector responses to
a GW signal.  It significantly simplifies the solution of the inverse problem
in the burst analysis and enables weakly modeled burst searches with the
polarization constraints.

The paper is organized as follows. 
Section~\ref{sec:overview} gives introduction into the coherent network analysis, 
required to introduce in section~\ref{sec:dualstream} the dual stream likelihood 
analysis and the polarization pattern. In sections~\ref{sec:netcon} we describe 
how it can be used to construct network regulators - the model independent constraints
used in the cWB analysis.  

\section{Overview}
\label{sec:overview}

A data from a network of $K$ detectors is presented as discrete series $x_k[i]$ in 
the most general time-frequency (TF) domain, where $k$ is the detector index in the network and $i$ 
is the data sampling (TF pixel) index. The real TF series $x_k[i]$ are obtained 
from the detector time series with the WDM transform~\cite{WDM}.
The data is conditioned to remove spectral features, such as violin, power and
mechanical lines~\cite{Tiwari:2015tkm}. 

A detector noise (assuming to be Gaussian) is described 
by the WDM power spectral density $S_k[i]$ estimated for every data sample.
Therefore, $S_k[i]$ is a TF series as well, which is convenient for
the characterization of a quasi-stationary noise typical for real detectors.   
The noise-scaled (whitened) data is defined as $w_k[i]=x_k[i]/\sqrt{S_k[i]}$.

The whitened TS series from all detectors are combined to obtain 
the energy TF maps $E[i]=\sum_k{w^2_k[i]}$, where $E[i]$ are maximized 
over all possible time-of-flight delays in the network. The energy
maps are used to identify TF areas (cluster $C$, ${i\in{C}}$) with the excess energy
above the baseline detector noise. The TF clusters, identified with an appropriate 
clustering algorithm, define the burst events,
which are analyzed to extract the signal waveform, 
polarization and sky location (inverse problem).

\subsection{Formulation of the inverse problem for bursts}
\label{sec:Iproblem}

The data vector ${\bf{x}}[i]=\{x_1[i],..,x_K[i]\}$ recorded by a network of GW detectors 
at the time of a gravitational-wave signal ${\bf{h}}[i]=[h_+[i],h_\times[i]]$
  with the source sky location at $\theta$ 
and $\phi$ is a superposition of the network response ${\bf{\cal{F}}}{\bf{h}}[i]$ and noise 
$\bf{n}[i]$: 
\begin{equation}
{\bf{x}}[i]={\bf{\cal{F}}}{\bf{h}}[i]+\bf{n}[i], 
\end{equation}
where the  $h_+$ and $h_\times$ are the amplitudes of the two GW polarization components 
and ${\bf{\cal{F}}}$ is the network antenna pattern matrix
\begin{equation}
\label{eq:dataMatrix}
  {\bf{\cal{F}}} =
  \left[ \begin{array}{cc} 
      F_{1+}(\theta,\phi) &  F_{1\times}(\theta,\phi) \\ 
      ... &  ... \\ 
      F_{K+}(\theta,\phi) &  F_{K\times}(\theta,\phi) \\ 
  \end{array} \right] \;.
\end{equation}
The antenna patterns often include a transformation by the polarization angle $\Psi$.
But this transformation is equivalent to a rotation of the wave frame where the
vector ${\bf{h}}$ is defined. The network response is $\Psi$-invariant 
and, therefore, the polarization angle can be included in the definition of ${\bf{h}}$.
 
To solve the inverse problem one should find the amplitudes 
of the GW polarization components $(h_+,h_{\times})$ and the sky coordinates $(\theta,\phi)$ 
from a coincident output of several GW detectors.  
Initially this problem was considered by Gursel and Tinto~\cite{GT} for a network of three
detectors.
A more solid statistical foundation of the problem was presented 
by Flanagan and Hughes~\cite{FH}, who considered a likelihood method for the estimation of the
signal parameters. They define the likelihood ratio
\begin{equation}
\label{eq:LIKE}
\Lambda({{\bf{x}},\Omega}) = \frac{p({\bf{x}}|{\bf{h}}(\Omega))}{p({\bf{x}}|0)}\;,
\end{equation}
where $\Omega$ is a parameter set describing the signal, the $p({\bf{x}}|0)$ is 
the joint probability that the data is only instrumental noise, and $p({\bf{x}}|{\bf{h}})$ 
is the joint probability that a GW signal ${\bf{h}}$ is present in the data ${\bf{x}}$.
The sample index $i$ is omitted to stress that ${i\in{C}}$, where $C$ is a collection 
of the TF pixels (cluster). 

The explicit form of the likelihood ratio is determined by the noise model 
$p({\bf{x}}|0)$ and by the signal model ${\bf{h}}(\Omega)$.
For un-modeled burst signals $\Omega=(h_+,h_{\times},\theta,\phi)$, which can be 
found by analytical or numerical variation of $\Lambda$. The advantage
of the likelihood method is that it allows introduction of the signal and noise models,
and can be applied to an arbitrary detector network.

\subsection{Un-constrained likelihood analysis}
\label{sec:standardL}
This section presents the solution of the inverse problem assuming that the burst parameter 
set $\Omega$ is not constrained by a source model and the noise of detectors in the network 
is quasi-stationary and Gaussian with the power spectral densities $S_1, ..., S_K$.
The noise-scaled data vector is than 
\begin{equation}
\label{eq:Dwhite}
{\bf{w}}[i]=\frac{x_1[i,\tau_1(\theta,\phi)]}{\sqrt{S_1[i]}}, ..., 
\frac{x_K[i,\tau_K(\theta,\phi)]}{\sqrt{S_K[i]}} \;,
\end{equation}
where the  detector amplitudes 
$x_k[i,\tau_k(\theta,\phi)]$ take into account the time-of-flight delays $\tau_k$ 
depending upon the source coordinates $\theta$ and $\phi$.
Respectively, the noise-scaled network response vector is 
\begin{equation}
\label{eq:XIh}
\bfxi[i]={\bf{F}}[i]{\bf{h}}[i],
\end{equation}
where ${\bf{F}}[i]$ is the noise-scaled antenna pattern matrix 
\begin{equation}
\label{eq:netMatrix}
  {\bf{F}}[i] =
  \left[ \begin{array}{cc} 
      \frac{F_{1+}(\theta,\phi)}{\sqrt{S_1[i]}} &  \frac{F_{1\times}(\theta,\phi)}{\sqrt{S_1[i]}} \\ 
      ... &  ... \\ 
      \frac{F_{K+}(\theta,\phi)}{\sqrt{S_K[i]}} &  \frac{F_{K\times}(\theta,\phi)}{\sqrt{S_K[i]}} \\ 
  \end{array} \right] 
\end{equation}
We also introduce the network matrix ${\bf{f}}$, which is obtained from
${\bf{F}}$ by the transformation to the Dominant Polarization Frame (DPF)
introduced by Klimenko et al~\cite{klimenko:2005kmrm}. 

The likelihood functional ${\cal L}$ is defined as twice the logarithm of 
the likelihood ratio $\Lambda$
\begin{equation}
\label{eq:like}
{\cal{L}}[{\bf h}] = 2({\bf{w}} \vert \bfxi)- (\bfxi \vert \bfxi) \, .
\end{equation}
where the inner products $({\bf{w}} \vert \bfxi)$ and 
$(\bfxi \vert \bfxi)$ are calculated over the TF cluster.

The solution for the GW waveforms ${\bf{h}}$
is found by variation of the likelihood functional {\cal L}[{\bf h}].
It is convenient to introduce the antenna pattern vectors ${\bf{f_+}}$
and ${\bf{f_\times}}$, which are simply the columns of the matrix ${\bf{f}}$
and satisfy the DPF conventions:
$({\bf{f_+}} \cdot {\bf{f_\times}})=0$ and  $|{\bf{f_\times}}|\le|{\bf{f_+}}|$. 
These two vectors define a network plane where the GW response  vector 
$\bfxi$  must be located.
The likelihood variation gives a system of linear equations
for the amplitudes $h_+[i]$ and $h_\times[i]$ (also defined in the DPF)
\begin{equation}
\label{eq:linear}
  \left[ \begin{array}{c} ({\bf{w}[i]} \cdot {\bf{e}}_+[i]) \\ 
      ({\bf{w}[i]} \cdot {\bf{e}}_\times[i]) \\ \end{array} \right] =
  \left[ \begin{array}{cc} 
      |{\bf{f_+}}[i]| &  0 \\ 
      0 & |{\bf{f_\times}}[i]| \\ 
  \end{array} \right] 
  \left[ \begin{array}{c} h_+[i] \\ h_\times[i] \\  \end{array} \right] \;.
\end{equation}
where ${\bf{e_+}}$ and ${\bf{e_\times}}$ are the unit vectors along 
${\bf{f_+}}$ and ${\bf{f_\times}}$ respectively.
Note,  the $2\times{2}$ matrix in Eq.~\ref{eq:linear} characterizes 
the network sensitivity to the two GW polarizations. 
The maximum likelihood ratio statistic is calculated by 
substituting the solutions into ${\cal L}[{\bf h}]$. The result can be written as
\begin{equation}
\label{eq:lMax}
L_{\mathrm{max}} = \sum_{i\in{C}} {\bf{w}}[i] P[i] {\bf{w}}^T[i] \;,
\end{equation}
where the matrix $P$ is the projection constructed from the components of the 
unit vectors ${\bf{e_+}}$ and ${\bf{e_\times}}$:
\begin{equation}
\label{eq:POP}
P_{nm}[i]=e_{n+}[i]e_{m+}[i]+e_{n\times}[i]e_{m\times}[i] \,.
\end{equation}
The kernel of the projection $P$ is the network plane defined by these two vectors. 
The null space of the projection $P$ defines the residual detector noise,
which is referred to as the null stream. 

\subsection{Reconstructed network response}
\label{sec:netresp}
The maximum likelihood ratio statistic $L_{\mathrm{max}}$  is a quadratic form
(see Eq.~\ref{eq:lMax}), which can be split into the incoherent $E_i$
and coherent $E_c$ parts
\begin{eqnarray}
\label{eq:Ei} 
E_i =  \sum_{i\in{C}} {\sum_{n}{w_n[i]P_{nn}[i]w_n[i]}} \;, \\
\label{eq:Ec} 
E_c =  \sum_{i\in{C}} {\sum_{n\neq{m}}{w_n[i]P_{nm}[i]w_m[i]}} \;.
\end{eqnarray}
These coherent statistics, together with the energy of the null stream
$E_n$, are widely used in the burst searches for the construction of
the event selection cuts.  For example, the network correlation 
coefficient~\cite{klimenko:2008kymm} 
\begin{equation}
\label{eq:cc}
c_c=E_c/(|E_c|+E_n) 
\end{equation}
provides a powerful event consistency test
to distinguish genuine GW events ($c_c\sim 1$) from spurious events ($c_c<<1$) 
produced by the detectors.
The statistic $E_c$ (coherent energy) is particularly important
because it depends on the cross-correlation terms between the detector
pairs. It is used for the construction of the burst detection statistic
\begin{equation}
\label{eq:etac}
\eta_c=(c_c E_c K/(K-1))^{1/2} \;,  
\end{equation}
which is an estimator of the 
network coherent signal-to-noise ratio 
for correlated GW signals recorded by different detectors. 

The coherent statistics are very beneficial for the burst analysis, 
provided they are correctly constructed to address the ``two-detector
paradox''~\cite{2Dparadox}. Namely, for any network of two detectors the cross
terms of the projection operator $P_{nm}[i]$ (Eq.~\ref{eq:POP}) 
are always equal to zero, or $E_c=0$. Clearly, for
two co-aligned detectors with the identical detector responses this is not true,
which constitutes the two-detector paradox. 

The origin of the ``two-detector paradox'' is the ambiguity of the 
projection operator. The likelihood $L_\mathrm{max}$ is
invariant with respect to the rotation in the network plane where
any two orthogonal unit vectors can be used for the construction of
the projection $P_{nm}[i]$. 
Therefore, we select two such unit vectors ${\bf{u}}[i]$ and ${\bf{v}}[i]$ 
that the likelihood component corresponding to the vector ${\bf{v}}[i]$ vanishes 
and the projection $P_{nm}({\bf{v}}[i])$ can be omitted.
The $L_\mathrm{max}$ and the coherent statistics are given by the projection
\begin{equation}
\label{eq:P(u)} 
P_{nm}({\bf{u}}[i])=u_n[i]u_m[i] \;,
\end{equation} 
which resolves the two-detector paradox. 
The vectors ${\bf{u}}[i]$ define the reconstructed network
response 
\begin{equation}
\label{eq:xi}
\bfxi_r[i] = ({\bf{w}}[i] \cdot {\bf{u}}[i]){\bf{u}}[i] \; ,
\end{equation}
which components are the un-constrained likelihood estimators  of 
the noise-scaled detector responses.


\section{Dual stream likelihood analysis}
\label{sec:dualstream}

As defined in section~\ref{sec:overview} a network noise-scaled data stream is ${\bf{w}}$. 
Additionally a quadrature data stream $\bf{\tilde{w}}$ is used. It can be obtained from 
the original detector data, which is phase-shifted by $-90^o$. 
The data ($\bf{{w}},\bf{\tilde{w}}$) defines the network dual data stream conveniently
provided by the WDM transform: applied to the detector time-series it generates both data streams. 
For the un-modeled reconstruction (see section~\ref{sec:standardL}) the analysis can be performed
individually for each data stream resulting in the likelihood statistics $L_{\text{max}}$ 
and $\tilde{L}_{\text{max}}$.  Formally, the quadrature data stream does not contain any
new information, nevertheless  $L_{\text{max}} \neq \tilde{L}_{\text{max}}$. This is 
because for a given time-frequency cluster the
quadrature counterparts may have different contributions both from the signal and noise. 
Therefore, the inclusion of the quadrature stream
can improve the collection of the signal energy 
and, respectively, improve the reconstruction. 
Also the dual data stream is required for the inclusion of the signal polarization models into the analysis.

\subsection{Phase transformation}
\label{sec:phasetrans}

Each dual data stream sample is presented by the data vectors 
${\bf{w}}[i]$ and ${\bf{\tilde{w}}}[i]$. We define a phase transformation
to calculate the amplitudes for an arbitrary phase shift $\lambda_i$:
\begin{eqnarray}
\label{eq:phase1}
\bf{w}'[i] = {\bf{w}}[i] \cos{\lambda_i} + \bf{\tilde{w}}[i] \sin{\lambda_i} \;, \\
\label{eq:phase2}
\bf{\tilde{w}'}[i] = {\bf{\tilde{w}}}[i] \cos{\lambda_i} - \bf{w}[i] \sin{\lambda_i} \;,
\end{eqnarray}
where the individual phase shift
$\lambda_i$ is applied to each data sample. 
In the likelihood functional the same transformation should be applied to the detector
responses $\bfxi[i]$ and $\tilde{\bfxi}[i]$.
The quadrature likelihood functionals ${\cal{L}}$ and $\tilde{\cal{L}}$ 
\begin{eqnarray}
\label{eq:UlikeD1}
{\cal{L}}[{\bf h}] = 2({\bf{w}} \vert \bfxi)- 
(\bfxi \vert \bfxi) \\
\label{eq:UlikeD2}
\tilde{\cal{L}}[{\bf h}] = 2({\tilde{\bf{w}}} \vert \tilde{\bfxi})- 
(\tilde{\bfxi} \vert \tilde{\bfxi})
\end{eqnarray}
vary as the phase transformation is applied, however the total likelihood 
${\cal{L}}_\circ = {\cal{L}} + \tilde{\cal{L}}$  
is the phase invariant.
There are several distinct phase transformations, two of which are considered below.

In the orthogonal phase transformation (OPT) the phase shift is selected such 
that the network responses  $\bfxi'[i]$ and $\tilde{\bfxi}'[i]$ in
the network plane become orthogonal to each other. The OPT pattern is used 
for calculation of dual stream coherent statistics in Section~\ref{index}.

The polarization phase transformation (PPT) is defined by the scalar products of
 the network response and the antenna pattern vectors
\begin{align}
\label{eq:pptC}
\cos{\lambda_{i}}  \propto (\bfxi[i] \cdot {\bf{e_+}}[i]) ,
~~\sin{\lambda_{i}} \propto (\tilde{\bfxi}[i] \cdot {\bf{e_+}}[i]) \;.
\end{align}

The purpose of the phase transformations is to obtain the signal polarization
patterns. Namely, the wave polarization is captured by the network as a distinct 
pattern of the GW responses in the network plane, which is revealed
when a particular phase transformation is applied.

\subsection{Polarization pattern}
\label{polar}

To describe the polarization state of a generic GW signal the
following parameterization of the wave is used:
\begin{align}
\label{eq:wavepar1}
\bfxi = ~~~{h \bf{F_+}}(\psi) + e H {\bf{F_\times}}(\psi) \;, \\ 
\label{eq:wavepar2}
\tilde\bfxi = -H {\bf{F_+}}(\psi) + e h {\bf{F_\times}}(\psi) \;, 
\end{align}
where the instantaneous parameters of the signal are: $h$ and $H$ are 
the strain amplitudes, $\psi$ is the polarization
angle and  $e$ is the wave ellipticity. Here and below in the text we omit the sample index $i$.
In general, these are the ad-hoc wave parameters, 
however, they can be related to the astrophysical wave parameters 
as described in section~\ref{pcon}. For this particular convention, 
the ${\bfxi}$ and $\tilde{\bfxi}$ are the $0^\circ$-phase and $-90^\circ$-phase 
network responses and the sign of $e$ defines the wave chirality or the sign of the quadruple product
$[{\bfxi} \times {\tilde\bfxi}] \cdot 
\left[{\bf{F_+}}(\psi) \times {\bf{F_\times}}(\psi) \right]$.
The antenna pattern vectors ${\bf{F_+}}(\psi)$ and ${\bf{F_\times}}(\psi)$
are related to the DPF vectors ${\bf{f_+}}$ and ${\bf{f_\times}}$
\begin{align}
\label{eq:F2DPF}
{\bf{F_+}}(\psi) = {\bf{f_+}}\cos(\gamma) - {\bf{f_\times}}\sin(\gamma) \;, \\ 
{\bf{F_\times}}(\psi) = {\bf{f_\times}}\cos(\gamma) + {\bf{f_+}}\sin(\gamma) \;, 
\end{align}
where $\gamma=\Psi-\psi$ and $\Psi$ is the DPF angle. 
The PPT pattern is obtained by application of the transformation~\ref{eq:pptC} to
the vectors $(\bfxi,\tilde\bfxi)$~\ref{eq:wavepar1}-\ref{eq:wavepar2}. The resulting 
PPT pattern is described by the following tree vectors
oriented along the ${\bf{f_+}}$ and ${\bf{f_\times}}$
\begin{align}
\label{eq:F2DPF1}
&{\bfxi_+} = {\bf{f_+}} h_\circ \beta_+(e,\gamma) ,& \\ 
\label{eq:F2DPF2}
&{\bfxi_\times} = -{\bf{f_\times}} h_\circ \frac{1-e^2}{2}\sin(2\gamma) \beta^{-1}_+(e,\gamma) ,& \\ 
\label{eq:F2DPF3}
&{\tilde\bfxi}_\times = {\bf{f_\times}} e h_\circ \beta^{-1}_+(e,\gamma) ,& 
\end{align}
where $h_\circ=\sqrt{h^2+H^2}$ is the wave amplitude and
\begin{eqnarray}
\label{eq:nu}
\beta_\pm(e,\gamma) = \frac{1}{\sqrt{2}}[1+e^2 \pm (1-e^2)\cos(2\gamma)]^{1/2} \;. 
\end{eqnarray}
The product $h_\circ {\beta_+(e,\gamma)}$ is the norm of $sin(\lambda_i)$ and
$cos(\lambda_i)$ in Equation~\ref{eq:pptC}.
 The vectors ${\bfxi_+}$ and ${\bfxi_\times}$
describe the $0^\circ$-phase network response, and the vector ${\tilde\bfxi_\times}$ describes
the $-90^\circ$-phase network response. By measuring these three vectors for each network data sample, 
the instantaneous signal parameters $h_o$, $e$ and $\psi$ can be determined. 

\begin{figure}[!bt]
 \begin{center} 
  \begin{tabular}{c}
 \includegraphics[width=80mm]{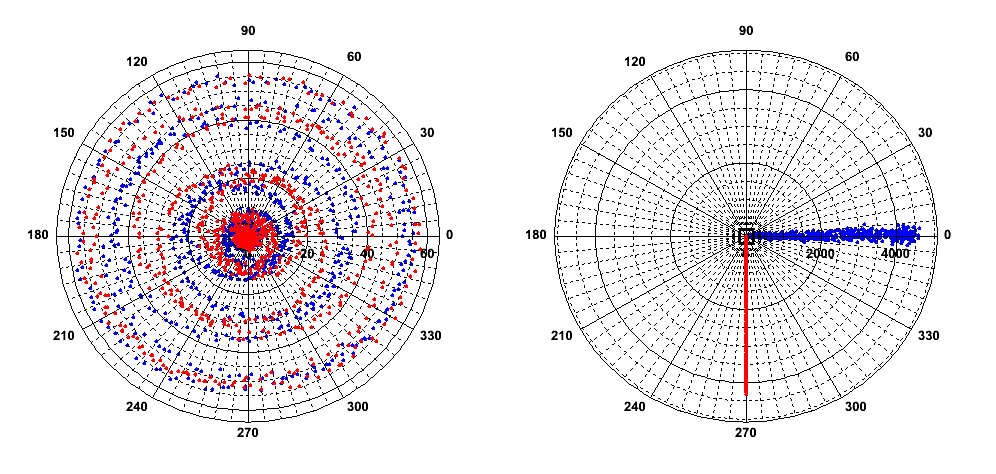} \\
 \includegraphics[width=80mm]{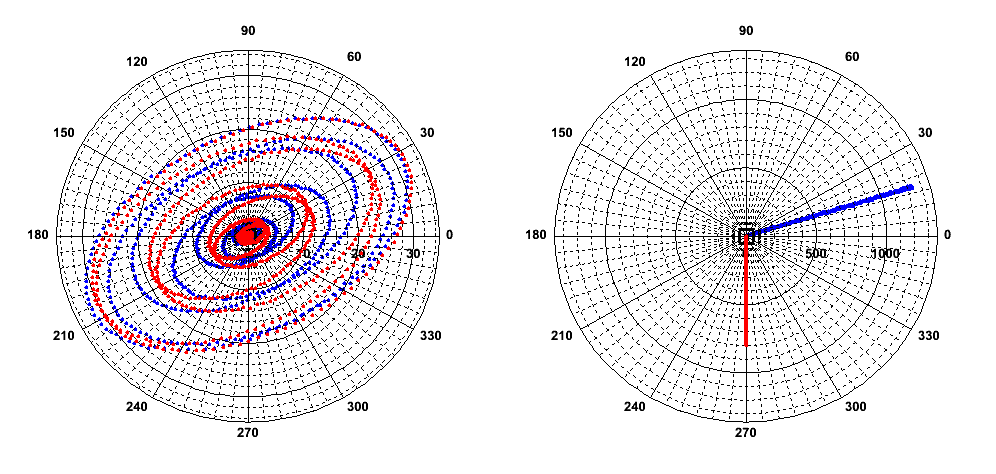} \\
 \includegraphics[width=80mm]{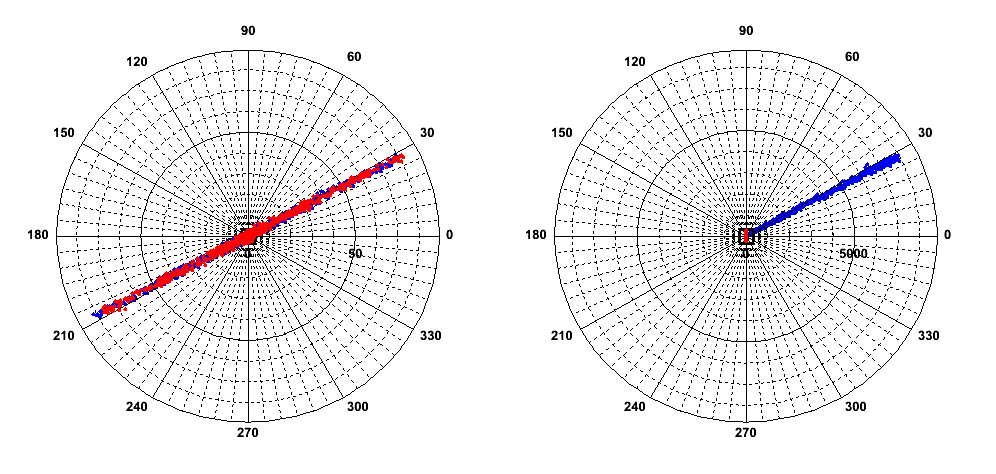} \\
 \includegraphics[width=80mm]{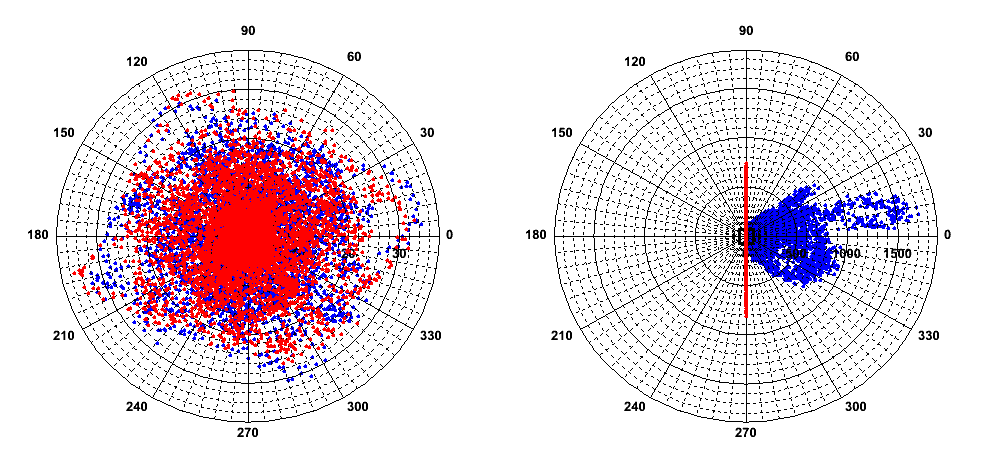} 
    \end{tabular}
 \end{center}\caption{\small{\textit{Polarization patterns (from top to bottom) for waves 
with circular, elliptical, linear and random polarization, before (left plots) and after
(right plots) the PPT is applied. Dots of two different colors
represent the orientation and amplitude of the $0^\circ$-phase (blue) and  -$90^\circ$-phase (red) 
response vectors originating at zero. The network antenna vectors form the coordinate frame 
with ${\bf{f}}_\times$ and  ${\bf{f}}_+$ pointing along the vertical and horizontal axis respectively. The patterns are calculated for a sky location where 
$\left|{\bf{f}}_\times\right| \approx \left|{\bf{f}}_+\right|$.}}}
\label{Fig:sgQ9circ}
\end{figure}

\begin{figure}[!ht]
 \begin{center} 
  \begin{tabular}{c}
 \includegraphics[width=80mm]{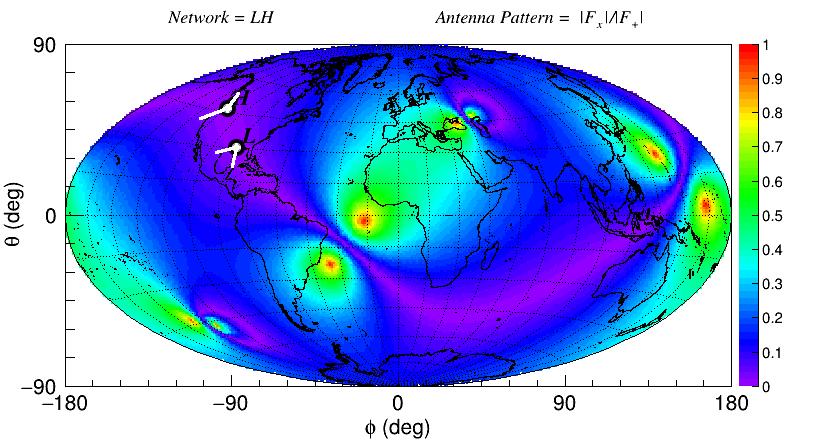} \\
 \includegraphics[width=80mm]{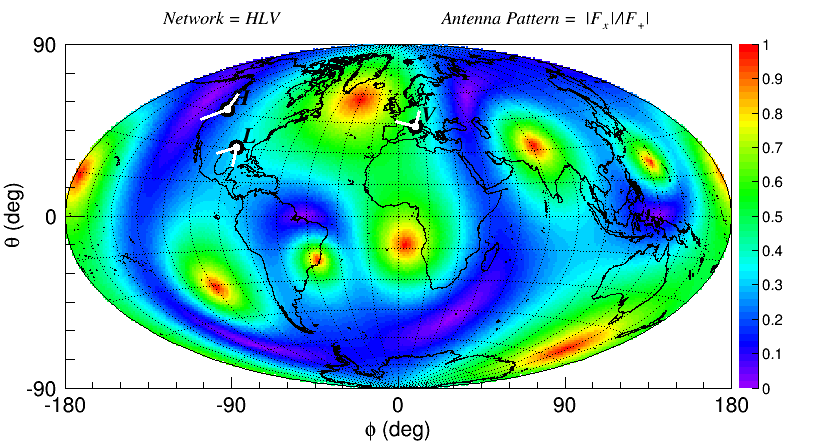} \\
 \includegraphics[width=80mm]{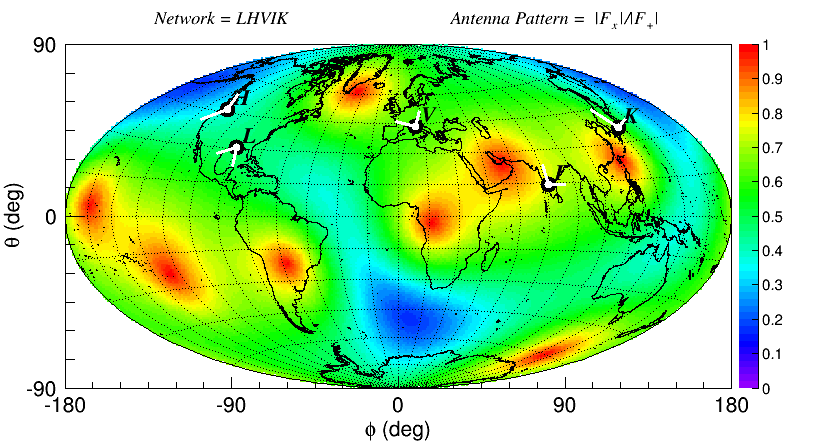} 
    \end{tabular}
 \end{center}\caption{\small{\textit{The distribution of $\alpha$ over the sky 
for Livingston-Hanford network (top), Livingston-Hanford-LIGOIndia (middle) 
and Livingston-Hanford-Virgo-Kagra-LIGOIndia (bottom). The detector site locations 
and the orientations of the arms are shown on the map. The LIGOIndia location is just
an example: there is no official site yet.}}}
\label{Fig:AP}
\end{figure}

For a given GW event, the collection of vectors \{${\bfxi_+}$,${\bfxi_\times}$,${\tilde\bfxi_\times}$\}
 describes its unique polarization pattern. Figure~\ref{Fig:sgQ9circ} shows 
the examples of the polarization patterns. 
However, this pattern can be significantly distorted by the
network. For example, the detector noise adds random vectors to
the GW responses and randomizes the polarization patterns for a weak GW signal.    
Also the measured polarization pattern strongly depends on the
network alignment factor: 
$\alpha = \left|{\bf{f}}_\times\right| / \left|{\bf{f}}_+\right|$.
For any practical network $\alpha < 1$, therefore the polarization pattern 
is always distorted (biased) by the network. The bias correction is 
straightforward, however, it becomes increasingly inaccurate when
$\alpha << 1$.  When $\alpha=0$ only the ${\bfxi_+}$ vector can be measured
regardless what is the GW polarization state. Namely, the original GW
polarization can not be reconstructed from such pattern of the network responses.
The network of LIGO detectors has
$\alpha << 1$ for a significant fraction of the sky (see top Figure~\ref{Fig:AP}).
\noindent 
Therefore, in most cases the polarization
state of a GW signal can not be measured. To improve the polarization coverage,
one has to increase the alignment factor by adding more detectors to 
the network with optimally oriented detector arms (see bottom Figure~\ref{Fig:AP}).
The full polarization coverage is achieved when $\alpha$ is close to
unity, which greatly improves and simplifies the reconstruction. In
this case, the polarization state of arbitrary GW signal can be un-ambiguously 
identified from the pattern of the network responses in the network plane. 
Also, a more complete polarization
coverage helps reconstruction of the sky coordinates and other
source parameters. 

\subsection{Polarization constraints}
\label{pcon}

The wave parameters $e[i]$ and $\psi[i]$ describe the polarization pattern. 
In some case they can be related to the astrophysical source parameters.
In this section we explicitly use the sample index $i$ to demonstrate 
that the event parameter may vary during its time-frequency evolution.
For example, for binary systems $e[i]$ are defined by the inclination
angle of the source and $\psi[i]$ are defined by the polarization angle.  
The parameters $e[i]$ and $\psi[i]$ can be constrained when
sources with a certain polarization  state are considered. 
For un-modeled signals all $e[i]$ and $\psi[i]$ are free parameters
or, in other  words, the wave polarization is random ($r$-waves). 
In this case, the solution for the network responses has been already
described in the previous section. By imposing constraints on $e[i]$ and 
$\psi[i]$, the $r$-waves can be divided into sub-classes with more definite 
polarization states. For example, most GW signals should produce patterns
with fixed chirality (all $e[i]>0$ or all $e[i]<0$).  Therefore, the vectors
${\tilde\bfxi_\times}$ can be constrained to have the same chirality
($\iota$-waves). A more narrow sub-class of $\iota$-waves are non-precessing  
binary systems, where the parameters $e[i]$ are related to the inclination angle 
of the source and therefore $e[i]=const$. 
The angles $\psi[i]$ define 
the orientation of the reconstructed response vectors in the network plane.
Assuming that the parameters $e[i]$ are free, the constraint
$\psi[i] = const$ describes a particular class of GW signals with the same direction
of the network response vectors ($\Psi$-waves).
The elliptical, linear and circular waves are defined when both angles $e[i]$ 
and $\psi[i]$ are constrained.
The constraints for the $\iota$-waves and $\Psi$-waves, and their combinations characterizing 
different polarization models are summarized in Table~\ref{tab:pmodels}. 
\begin{table}[h]
\centering
\begin{tabular}{|c|c|c|c|}\hline
$e$      & $\psi$        & pattern  & polarization \\
constraint & constraint & constraint & state \\
\hline
-                   & -               & - & r-waves \\
$sign(e[i])=const$  & -         & - & $\iota$-waves \\
-                   & $\psi[i]=const$ & - & $\psi$-waves \\
$e[i]=const$        & $\psi[i]=const$ & - & elliptical \\
$e[i]=0$            & $\psi[i]=const$ & - & linear \\
 -                  & - & $\tilde\bfxi_\times=0$ & loose linear \\
$e[i]=\pm{1}$       & -               & - & circular \\
 -                  & -   & ${\bfxi_\times}=0$ & loose circular \\
\hline
\end{tabular}
\caption{\small{\textit{The constraints on $e$ (first column), $\psi$ (second column)
and the pattern vector (third column). 
The corresponding polarization states are shown in the last column.}}}
\label{tab:pmodels}
\end{table}
The simplest solution is for the waves with the circular polarization:
$e[i]=\pm{1}$. A less strict (loose) circular polarization constraint is when
$\bfxi_\times=0$. In this case the network responses are defined by
the vectors $\bfxi_+$ and $\tilde\bfxi_\times$ and the condition
$e[i]=\pm{1}$ is not enforced. 
For linear waves $e[i]=0$ 
and all $0^\circ$-phase response vectors are co-aligned, or $\psi[i]=const$. 
Respectively, a less strict (loose) linear polarization constraint
is defined by the condition $\tilde\bfxi_\times=0$ when the condition
$\psi[i]=const$ is not enforced.
The polarization constraints can be used to construct weakly modeled burst searches
targeting broad classes of GW transients. The $\iota$-wave constraint can be applied to 
any rotating source. The elliptical, circular and the $\psi$-wave constraints 
can be used to search for compact binary sources with different spin configurations.

\subsection{Likelihood solutions}
\label{DSLS}

The solution for the wave parameters  $ h_\circ$, $e$ and $\psi$, and
hence, the waveforms $\bfxi$ and $\tilde\bfxi$, can be obtained by maximizing 
the likelihood functional  in Equations~\ref{eq:UlikeD1}-\ref{eq:UlikeD2}.
For un-constraints case when all the wave parameters are free, it is 
straightforward to show that the solutions for the network responses 
are given by the projections of the data vectors 
($\bf{{w}}$,$\bf{\tilde{w}}$) on the network plane. 
As described above the un-modeled burst analysis can be constrained to search
for GW signals with various polarization states. In general case, the constrained 
likelihood problem is hard to solve analytically and the numerical 
solutions are computationally prohibitive. To solve this problem,  we apply 
the phase transformation in Equation~\ref{eq:pptC} to the data vectors 
${\bf{w}}$ and $\tilde{\bf{w}}$.
This transformation reveals the underlying polarization pattern 
\{${\bf{w}_+}$,${\bf{w}_\times}$,${\tilde{\bf{w}}_\times}$\} smeared by the detector noise.
The detector responses can be reconstructed directly from this pattern. 
The solutions for different polarization states can be obtained by imposing the 
polarization constraints in Table~\ref{tab:pmodels}. 
As follows from Equations~\ref{eq:F2DPF1}-\ref{eq:F2DPF3}, 
for linear ($e=0$) and circular ($e\pm{1}$) waves the components 
$\tilde\bfxi_{\times}=0$ and $\bfxi_{\times}=0$ respectively. 
Therefore, the reconstructed responses for the loose linear polarization constraint are
($\bfxi_+={\bf{w}}_+,\bfxi_{\times}={\bf{w}}_\times,\tilde\bfxi_{\times}=0$)
and for the loose circular polarization constraint they are  
($\bfxi_{+}={\bf{w}}_+,\bfxi_{\times}=0,\tilde\bfxi_{\times}=\tilde{\bf{w}}_\times$).
The solution for linear waves is 
($\bfxi_{+}={\bf{p}}_+,\bfxi_{\times}={\bf{p}}_\times,\tilde\bfxi_{\times}=0$)
where ${\bf{p}}$ are the projections of ${\bf{w}_+}+{\bf{w}_\times}$ on their average vector.
The analytic solutions for the other polarization constraints are straightforward 
to find and we do present them here. Such significant simplification
of the inverse problem is possible due to the polarization transformations 
introduced in this paper. It enables rapid searches over the entire sky and 
reconstruction of source coordinates in real time. 

\subsection{Sky localization}
\begin{figure}[!ht]
 \begin{center} 
  \begin{tabular}{c}
 \includegraphics[width=80mm]{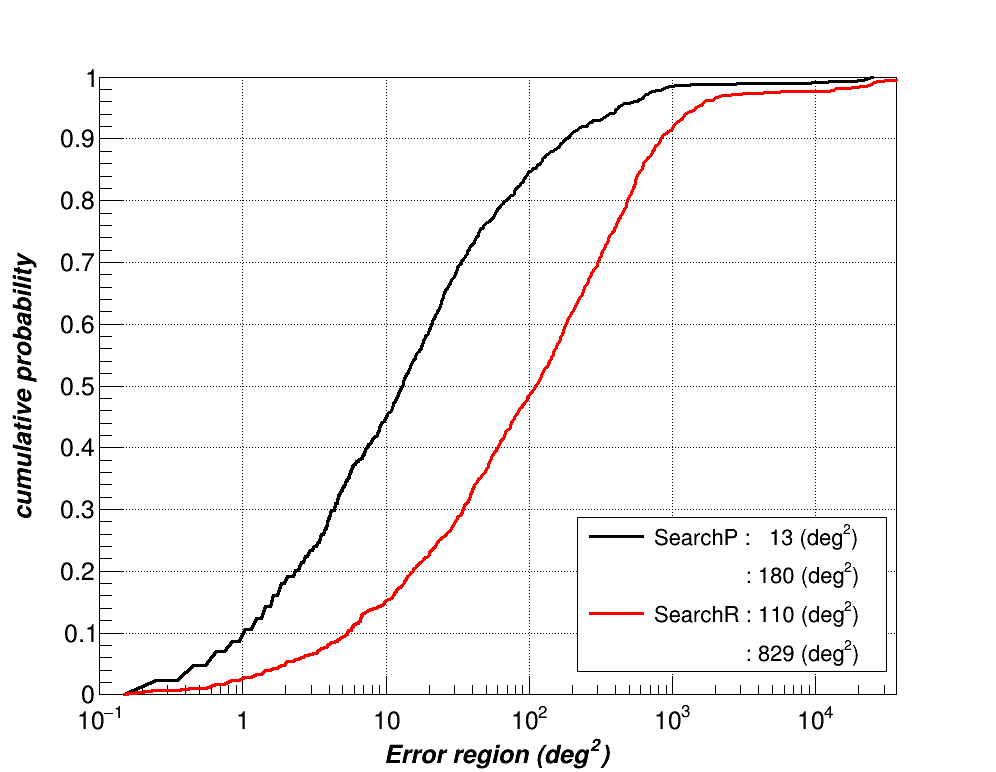}
    \end{tabular}
 \end{center}\caption{\small{\textit{Fraction of sources (vertical axis) reconstructed
by advanced Livingston-Hanford-Virgo network (at designed sensitivity) within the error
region in square degrees (horizontal axis: the legend shows the median search area) 
for a simulated population of binary black holes: uniform 
in volume distribution with component masses between 
15 and 25 solar mass and spin parameter between 0 and 0.9}}}
\label{Fig:skyLoc}
\end{figure}

As described in Section~\ref{sec:standardL} the maximum likelihood and 
other coherent statistics are functions of the sky coordinates $\theta$ and $\phi$.
They are sensitive to the arrival time of a GW signal at the detector sites and
can be used for the source localization. The reconstructed source location
is defined at the maximum of the likelihood statistic $L_{\mathrm{max}}(\theta,\phi)$ 
or the sky statistic
\begin{equation}
\label{eq:skystat}
L_{\mathrm{sky}}(\theta,\phi) = c_c(\theta,\phi) L_{\mathrm{max}}(\theta,\phi).
\end{equation}
The $L_{\mathrm{sky}}$ statistic has better performance than $L_{\mathrm{max}}$ 
for networks with two detectors and both statistics have comparable performance
for larger networks. The probability distribution over the sky is calculated as
\begin{equation}
\label{eq:skystat}
P_{\mathrm{sky}}(\theta,\phi) \propto  
\left(\left|{\bf{f}}_+\right|^2+\left|{\bf{f}}_\times\right|^2\right)^n 
\exp{\left[\frac{S-S_\circ}{2\sigma^2_s}\right]} \;,
\end{equation}
where $S$ is either $L_{\mathrm{max}}$ or $L_{\mathrm{sky}}$, $S_\circ$ is
the maximum of $S$ in the sky and $\sigma_s$ is the scaling parameter close to unity.
The parameter $n=2$ invokes the antenna pattern 
prior function used for networks with two detectors. 
For $n=0$ the prior is not used. The scaling parameter
$\sigma_s$ could vary depending on the network and used to calibrate the 
probability  $P_{\mathrm{sky}}$, so it correctly represents the fraction of sources 
found in a given error region. Figure~\ref{Fig:skyLoc} shows the sky localization
performance of the advanced Livingston-Hanford-Virgo network for a population of 
simulated signals expected from mergers of compact binary sources.
It is characterized by the median search area defined as the size of the error region 
in the sky containing $50\%$ of sources.
Also Figure~\ref{Fig:skyLoc} shows that the $\psi$-wave constraint
significantly improves the source localization.
This is an expected improvement for sky localization constrained by the source 
models~\cite{Singer:2014qca}. Of course, any modeled sky localization can be biased 
when the model does not accurately match the observation. However, the $\psi$-wave 
constraint uses a very general assumption about the compact binary sources and
no significant bias is expected.  
 
\section{Network constraints}
\label{sec:netcon}

The polarization constraints should be distinguished from 
the network constraints (or regulators), which give
a model-independent way to constrain the wave
parameters $h_o[i]$, $\psi[i]$ and $e[i]$.
Main purpose of the regulators is to eliminate unlikely solutions
of the likelihood functional and, 
therefore, reduce the false alarm rates due to the
instrumental and enviromental artifacts in the data. 
The first network regulators were introduced by 
Klimenko et al~\cite{klimenko:2005kmrm} to utilize the
network properties in the likelihood analysis.  Depending on the
configuration, detector noise and sky location, the detector
network may have much lower sensitivity to the second GW component:
$|{\bf{f_\times}}|<<|{\bf{f_+}}|$. In this case most of the network
SNR is produced by the ${\bf{f_+}}$ response (see Eq.~\ref{eq:linear}.)
The ${\bf{f_\times}}$ network response
is likely to yield low SNR and therefore may not be
reconstructed from the noisy data.  Such a priori knowledge can be
used in the analysis to constrain the likelihood solutions and reduce
the number of free parameters in the wave model.

\subsection{Network and event index}
\label{index}

The weight of each detector in the network is defined by its noise-scaled 
response (Eq.~\ref{eq:XIh}).
Depending on the spectral characteristics of the detector noise and the source 
sky location, the detector can be a key player in the network or just a spectator. 
The detector role varies from event to event and with time, depending on the
run conditions. The quality of the network depends on how many detectors can
contribute to the measurement. It is characterized by the network index      
\begin{equation}
\label{eq:netind}
I_n = \frac{ \left|{\bf{f}}_+\right|^2+\left|{\bf{f}}_\times\right|^2 }
{\left|{\bf{f}}_+\right|^2\nu({\bf{e_+}})+\left|{\bf{f}}_\times\right|^2\nu({\bf{e_\times}})} \;,
\end{equation}
where $\nu({\bf{e}})=\sum_k{e^4_{k}}$ for any unit vector ${\bf{e}}$.  
The network index is distributed between 1 and K representing the
effective number of detectors available for the measurement. It is useful
to introduce also the event index 
\begin{equation}
\label{eq:evntind}
I_e = \frac{ |\bfxi'|^2+|\tilde\bfxi'|^2 }
{|\bfxi'|^2\nu({\bf{u}})+|\tilde\bfxi'|^2\nu({\bf{v}})} \;,
\end{equation}
where the unit vectors ${\bf{u}}$ and ${\bf{v}}$ are 
along the OPT vectors $\bfxi'$ and $\tilde\bfxi'$ respectively 
(see Section~\ref{sec:phasetrans}).
The event index is representing the effective number of coincident detectors participating 
in the measurement. Usually, a low value of $I_e$ or a significant difference between 
$I_n$ and $I_e$ is an indication of a spurious event produced by the detector noise. 

Note, for calculation of the event index and the other coherent statistics, the reconstructed
responses should be transformed to the OPT pattern  \{$\bfxi',\tilde\bfxi'$\},
where the vectors  ${\bf{u}}$ and ${\bf{v}}$ are orthogonal. They define the projection 
operators $P_{nm}({\bf{u}})$ and $P_{nm}({\bf{v}})$ respectively (see Equation~\ref{eq:P(u)}).
The coherent statistics \ref{eq:Ei}-\ref{eq:Ec} are calculated individually 
for the $0^\circ$-phase and $-90^\circ$-phase data and combined together. 

\subsection{Regulators}

As prescribed by the un-constrained likelihood analysis, the
orientation of the reconstructed response $\bfxi$ is always
along the unit vector ${\bf{u}}$ (see Eq.~\ref{eq:xi}.)  
However, when $|{\bf{f_\times}}|=0$, which
is the case for detectors with co-aligned arms, the true network response must
be pointing along the vector ${\bf{f}}_+$.  Therefore, instead of the
vector ${\bf{u}}$,  the unity vector along ${\bf{f}}_+$ must be selected for the
projection. This constitutes the hard
regulator, which constrains the
likelihood analysis to ignore the $\times$-response of the network.
This and several other regulators  have been used 
to analyze data collected by the initial LIGO and Virgo detectors.

\begin{figure}[!hbt]
 \begin{center} 
  \begin{tabular}{c}
 \includegraphics[width=80mm]{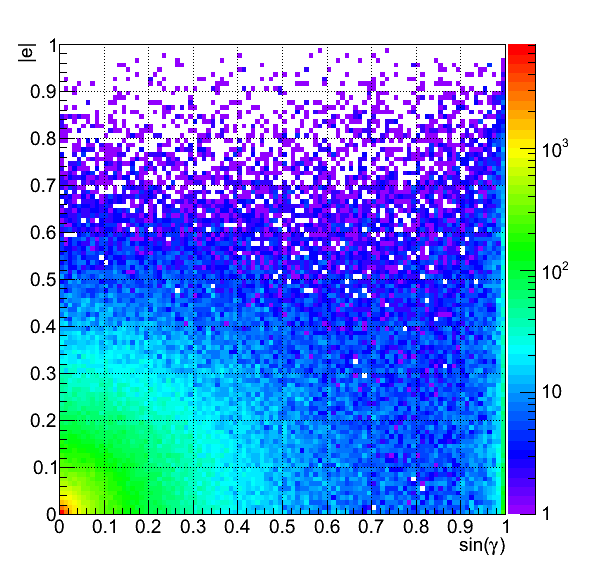} \\
 \includegraphics[width=80mm]{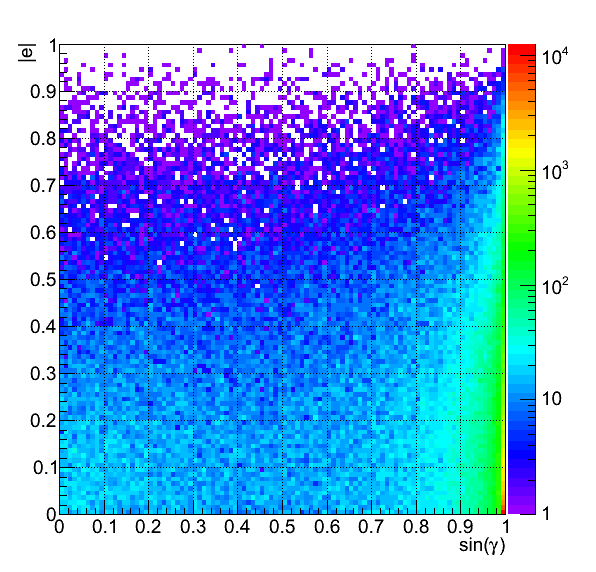} 
    \end{tabular}
 \end{center}\caption{\small{\textit{Distribution of $e$ and $sin(\gamma)$ for 
Livingston-Hanford network obtained 
from the simulation of a single-pixel events uniformly distributed over the sky. The top plot
is for Gaussian noise and the bottom plot is for signal with the random polarization.  }}}
\label{Fig:Greg}
\end{figure}

Given a network of detectors, in some cases it is possible to predict 
the distributions of the wave parameters and anticipated network responses to 
a generic GW signal. The polarization transformation significantly simplifies 
the construction of regulators. 
After substituting the left side of the Equations~\ref{eq:F2DPF1}-\ref{eq:F2DPF3} 
with the data pattern vectors \{${\bf{w}_+}$,${\bf{w}_\times}$,${\tilde{\bf{w}}_\times}$\} 
we obtain the following identities
\begin{align}
\label{eq:IDNx1}
 \alpha^2|{\bf{w}_+}|^2  
&=  |{\bf{f_\times}}|^2 h^2_\circ \beta^2_+(e,\gamma) , \\ 
\label{eq:IDNx2}
 |{\bf{w}_\times}|^2+|{\tilde{\bf{w}}_\times}|^2  
&=  |{\bf{f_\times}}|^2 h^2_\circ \beta^2_-(e,\gamma) , \\ 
\label{eq:IDNx3}
-\alpha\left({\bf{w}_\times}\cdot{\bf{e_\times}}\right) |{\bf{w}_+}|   
&= |{\bf{f_\times}}|^2 h^2_\circ \frac{1-e^2}{2}\sin(2\gamma) , \\ 
\label{eq:IDNx4}
\alpha\left({\tilde{\bf{w}}_\times}\cdot{\bf{e_\times}}\right) |{\bf{w}_+}|   
&= |{\bf{f_\times}}|^2 h^2_\circ e ,
\end{align}
that can be solved for $e$ and $sin(\gamma)$. 
As prescribed by Equations~\ref{eq:F2DPF2}-\ref{eq:F2DPF3},
the responses $\bfxi_\times$ and  $\tilde\bfxi_\times$ vanish 
when $sin(\gamma) \to 0$ and $e\to 0$ 
respectively. Figures~\ref{Fig:Greg} show the distributions of the reconstructed 
$e$ and $sin(\gamma)$ for noise and signal, and the Livingston-Hanford network. 
Unlike the signal, the noise is clustering at the low values of $e$ and $sin(\gamma)$. 
The noisy data can be identified by the regulator
\begin{equation}
\label{eq:regG}
\Gamma = \sqrt{e^2 + sin^2(\gamma)}
\end{equation}
when $\Gamma$ is below some threshold $\Gamma_o$. For appropriately selected $\Gamma_o$,
 the regulator identifies data pixels with the marginal signal 
components $\bfxi_\times$ and  $\tilde\bfxi_\times$ and zeroes them.
The regulated responses ($\bfxi_{+}={\bf{w}}_+,\bfxi_{\times}=0,\tilde\bfxi_{\times}=0$)
are biased: a small fraction of events can be miss-reconstructed and excluded from the
analysis. Despite this relatively small (and controlled) loss, the regulator otherwise
is very efficient in reducing the false alarm rates, with a typical reduction factor
of $\sim 10^{-6}$. It entirely eliminates the
single detector FAR and significantly suppresses FAR from the accidental coincident
events produced by the detector pairs. To further reduce the double coincidence FAR, 
we introduce the second regulator, which utilizes the network and the event indexes
\begin{equation}
\label{eq:regD}
\Delta = I^{-1}_e - \alpha|{\nu(\bf{e_+}})-\nu({\bf{e_\times}})|.
\end{equation}
The condition $\Delta>0.5$ is used to identify the situation when two or less 
detectors are used in the measurement.
In this case the reconstructed responses are constrained to be
($\bfxi_{+}={\bf{w}}_+,\bfxi_{\times}=0,\tilde\bfxi_{\times}=\tilde{\bf{w}}_\times$).
Both regulators can be used to constrain the detector networks
when either the network alignment coverage is insufficient,
or the effective number of detectors is less than 2. 

\section{Conclusion}
\label{conclusions}

In the paper we present the improved analytic framework of the cWB analysis algorithm. 
The objective of this analysis is the detection and reconstruction of 
un-modeled GW transients. It is achieved by solving the burst inverse 
problem - reconstruction of the signal waveforms, wave polarization and 
the sky coordinates of the source. The reconstruction is performed by using
the likelihood formalism with the signal waveforms as free parameters.
The waveforms can be described with the wave parameters and constrained,
which enable a range of weakly modeled burst searches. The likelihood analysis
yields a number of detection statistics used for the ranking of detected events
(the network coherent SNR), the rejection of background events (the network correlation
coefficient) and for the sky localization.    

The novelty of the paper is in the introduction of the polarization patterns.
By imposing a simple phase transformation to the network data, a characteristic
pattern emerges revealing the polarization state of an arbitrary GW signal.
This unique signature of the signal can be measured independent from the other 
source parameters. 
The polarization transformation significantly simplifies the solution of the inverse
problem: the detector responses can be reconstructed directly from the pattern.
The polarization constraints can be imposed, which enable weakly
modeled burst searches. The reconstruction is computationally efficient 
allowing for rapid searches over the entire sky and the
reconstruction of source coordinates in real time with a few minutes latency. 
We also identify factors limiting reconstruction and how the polarization 
measurements are affected by the network.
A simple metric (network alignment factor $\alpha$) determines 
the network ability to capture polarizations. 
The network of LIGO and Virgo detectors has a low alignment coverage for 
a significant fraction of the sky. Therefore, in most cases, the polarization state 
of a weak GW signal can not be measured. Adding Kagra 
and LIGO-India detectors to the advanced network will significantly improve 
the alignment coverage and, hence, the reconstruction of the signal parameters.

\section{Acknowledgements }

We are thankful to the  National Science Foundation for support 
under grants PHY 1205512 and PHY 1505308. This document has been assigned 
LIGO Laboratory document number P1500206.



\end{document}